# Impact of H.E.S.S. Lidar profiles on Crab Nebula data

*Justine Devin*[*1], *Johan Bregeon*[1], *Georges Vasileiadis*[1] *and Yves Gallant*[1]

[1]Laboratoire Univers et Particules de Montpellier, Université de Montpellier, CNRS/IN2P3, 34095 Montpellier, France

**Abstract.** The H.E.S.S. experiment in Namibia is a high-energy gamma-ray telescope sensitive in the energy range from 30 GeV to a several tens of TeV, that uses the atmospheric Cherenkov technique to detect showers developed within the atmosphere. The elastic lidar, installed on the H.E.S.S. site, allows to reduce the systematic errors related to the atmospheric composition uncertainties thanks to the estimation of the extinction profile for the Cherenkov light (300–650 nm). The latter has a direct impact on the reconstructed parameters, such as the photon energy and the source flux. In this paper we report on physics results obtained on the Crab Nebula spectrum using the lidar profiles obtained at the H.E.S.S. site.

## 1 Introduction

The H.E.S.S. experiment (High Energy Stereoscopic System) consists of five imaging Cherenkov telescopes situated in the Namibia Khomas Highland desert (1800 m above sea level) [1]. Its main objective is the study of Galactic and extragalactic sources in the energy range of ~ 30 GeV to several tens of TeV coupled to a high flux sensitivity (~ 1% Crab units). The detection technique of Cherenkov light from showers induced in the atmosphere by the incoming gamma ray demonstrates by itself the importance of knowing any variation on the atmosphere transparency, which is necessary to estimate the amount of light loss. For a shower induced by a photon of a given energy generating Cherenkov light that reaches the telescopes, the total charge measured by the cameras will be different if it propagates into a more transparent (or more opaque) atmosphere than the expected one. This directly impacts the reconstructed parameters, such as the photon energy and the source flux and can gives rise to an inaccurate reconstructed spectrum when the atmosphere transparency deviates significantly from the standard model. In this paper, we first present the lidar instrument at the H.E.S.S. site and the implementation of the lidar absorption profiles in the simulation. We then derive the instrument response functions associated to Crab Nebula observations, using the lidar absorption profiles (called lidar profiles) and the standard model (called standard profile) in order to compare them. We also quantify the impact on the reconstructed spectrum and in particular, we study the normalization dispersion which should in principle be reduced when using lidar data instead of the model currently used by the H.E.S.S. collaboration.

## 2 H.E.S.S. lidar and data implementation

### *2.1 The lidar instrument*

Conceived and constructed in 1997, the lidar was transferred to the H.E.S.S. site in 2004 and data collection began two years later. In order not to disturb the signal measured by the H.E.S.S. cameras, the lidar is located at ~ 850 m from the telescopes, always pointing in the same direction (zenith angle of 15° and 25° in the West direction), and takes data before H.E.S.S. observation runs. The lidar technique is the same as that of the radar, transposed in the optical wavelengths: the instrument measures the intensity of a laser induced backscattered signal as a function of the travel time of the photon return trip to the collecting area. The instrument uses a Nd: YAG type laser and is equipped with two cavities generating two harmonics at 355 nm and 532 nm. The reflected light is collected by a primary mirror and a secondary mirror, respectively with a diameter of 60 cm and 8 cm and a focal length of 102 cm and 10 cm. The reflectivity of the mirrors is 80% between 300 and 600 nm. A dichroic filter, located at the focal plane, separates the 355 and 532 nm components and photo-multiplicator tubes are used to measure the return signal. Capable of detecting signals up to ~ 15 km, the lidar sends 1200 light pulses over a period of 3 minutes before each H.E.S.S. observation run. The acquisition is then recorded on a server, connected to that of the H.E.S.S. data acquisition. A detailed technical description of the H.E.S.S. lidar and its performance can be found in [2].

### *2.2 Analysis implementation*

The most commonly used methods for deriving the backscatter coefficient β from a lidar are based on the inversion methods of Klett [3] and Fernald [4]. First, the background signal (electronic pedestal, sky background optical noise, etc.) is substracted from the raw signal, and corresponds to the average value of the signal between 20 and 25 km where no return signal is expected. The signal is then corrected for inclination (15° zenith angle) and the overlap function of the lidar. The inversion algorithm of the lidar equation requires a reference altitude $R_0$, defined such that only Rayleigh scattering is present. For the profiles used in this study, $R_0$ has been defined such that the ratio of Rayleigh scattering to noise is greater than 5. Since the use of an elastic lidar requires a hypothesis on the type of aerosols, we have chosen a lidar ratio $S_p = 50$, which seems appropriate for desert type of climates [5]. The inversion algorithm then gives, for each altitude interval, the value of the backscatter coefficient $\beta$ and, using the $S_p$ mentioned above, the absorption coefficient $\alpha$. The extinction profile derived from lidar data at 532 nm is used to scale the full light transmission table used in the simulations, for all wavelengths from 200 to 700 nm. Typical lidar profiles for two distinct atmospheric conditions are shown in Figure 1.

* Corresponding author: jdevin.phys@gmail.com

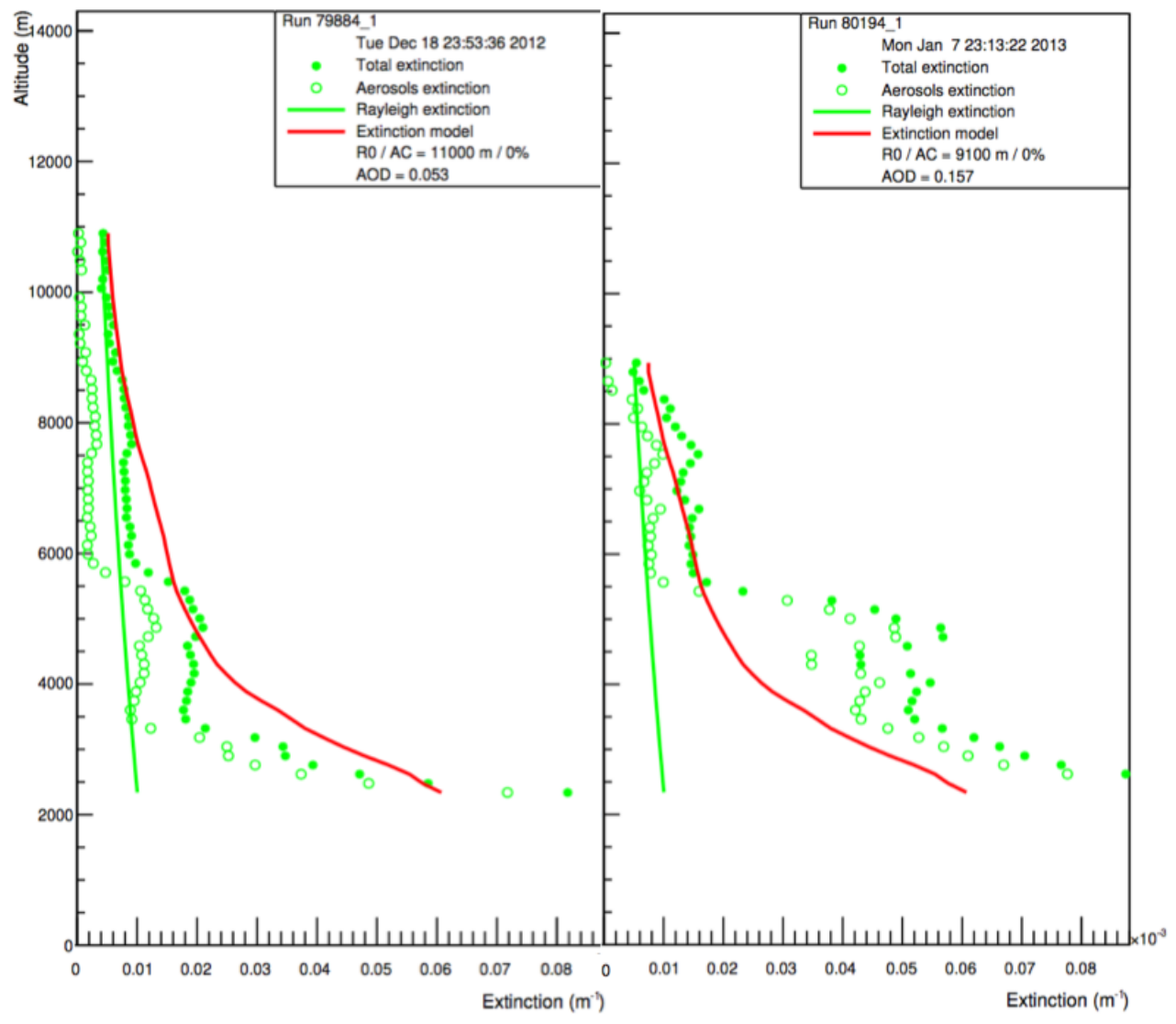

**Fig. 1.** Atmospheric profiles measured at 532 nm for two different observation runs. The red curve represents the standard model used by the H.E.S.S. collaboration and the extinction due to Rayleigh scattering is represented by the green line. Solid and empty green dots correspond respectively to the total extinction of the two scatterings (Rayleigh and Mie) and to the Mie scattering only.

In the classical H.E.S.S. analysis chain, the Instrument Response Functions (IRFs) are simulated for different zenith angles, optical efficiencies, and so on. The IRFs are then interpolated to cover the entire parameter space corresponding to all H.E.S.S. observations. A project was recently undertaken within the H.E.S.S. collaboration to simulate the IRFs using the real observation conditions of each run [6]. These simulations, called Run-Wise Simulations (RWS), aim to obtain the most realistic IRFs associated with each observation. This new simulation chain permits to achieve a greater accuracy of results, an improvement in our knowledge of the H.E.S.S. PSF, and leads to a reduction of the systematic errors. Since the use of a lidar aims to reduce the systematic uncertainties, we use this framework to be able to isolate the impact of the atmospheric profiles on the data. We first compare run by run the IRFs (effective areas and energy biases) obtained with the two atmospheric profiles (standard and lidar), before quantifying the impact on the spectral reconstruction. As the lidar measures the instantaneous composition of the atmosphere, the Crab Nebula spectrum must be better reconstructed with the lidar profiles. Thus, we also study the normalization dispersion which should in principle be reduced when using lidar data.

## 3 Crab Nebula spectral studies

We used data from two distinct periods (in 2012 and 2013), for which we have exploitable lidar profiles (for a total of 22 observation runs). The atmospheric conditions and the number of participating telescopes are different in the dataset from 2012 and from 2013. Data taken in 2012 include all 4 HESS-I telescopes and the large HESS-II telescope in the central trigger, and have mostly better atmospheric conditions, while data taken in 2013 include only the 4 HESS-I telescopes in the central trigger and have generally a strong presence of clouds and aerosols. Part of these runs did not pass the standard quality selection criteria defined in the H.E.S.S. collaboration. The total extinction of the atmosphere is due to both Rayleigh and Mie scattering. The optical depth (OD) of the atmosphere corresponds to the total extinction integrated over altitude. Figure 2 shows the OD values, at 532 nm, as a function of the altitude for the 22 runs used in this analysis. The most transparent and opaque atmosphere are respectively found for the runs 79884 and 80194.

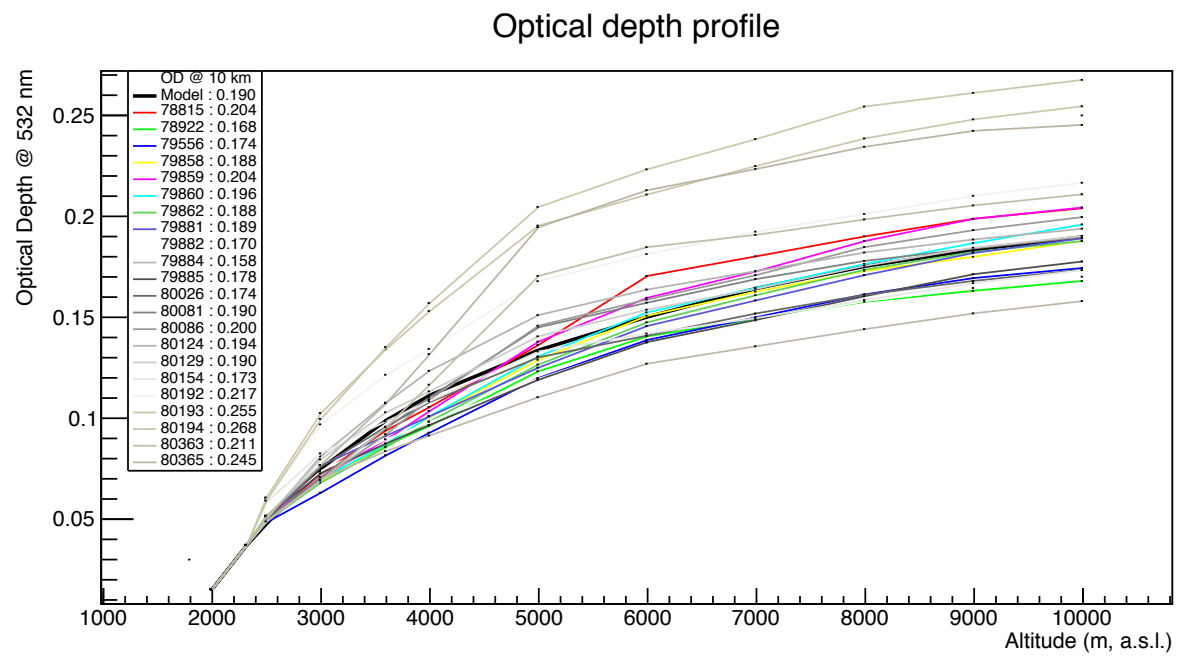

**Fig. 2.** Optical depth (OD) values at 532 nm as a function of the altitude for the 22 runs used in this analysis. The black curve represents the OD of the standard model. The OD values at 10 km are given on the top left corner.

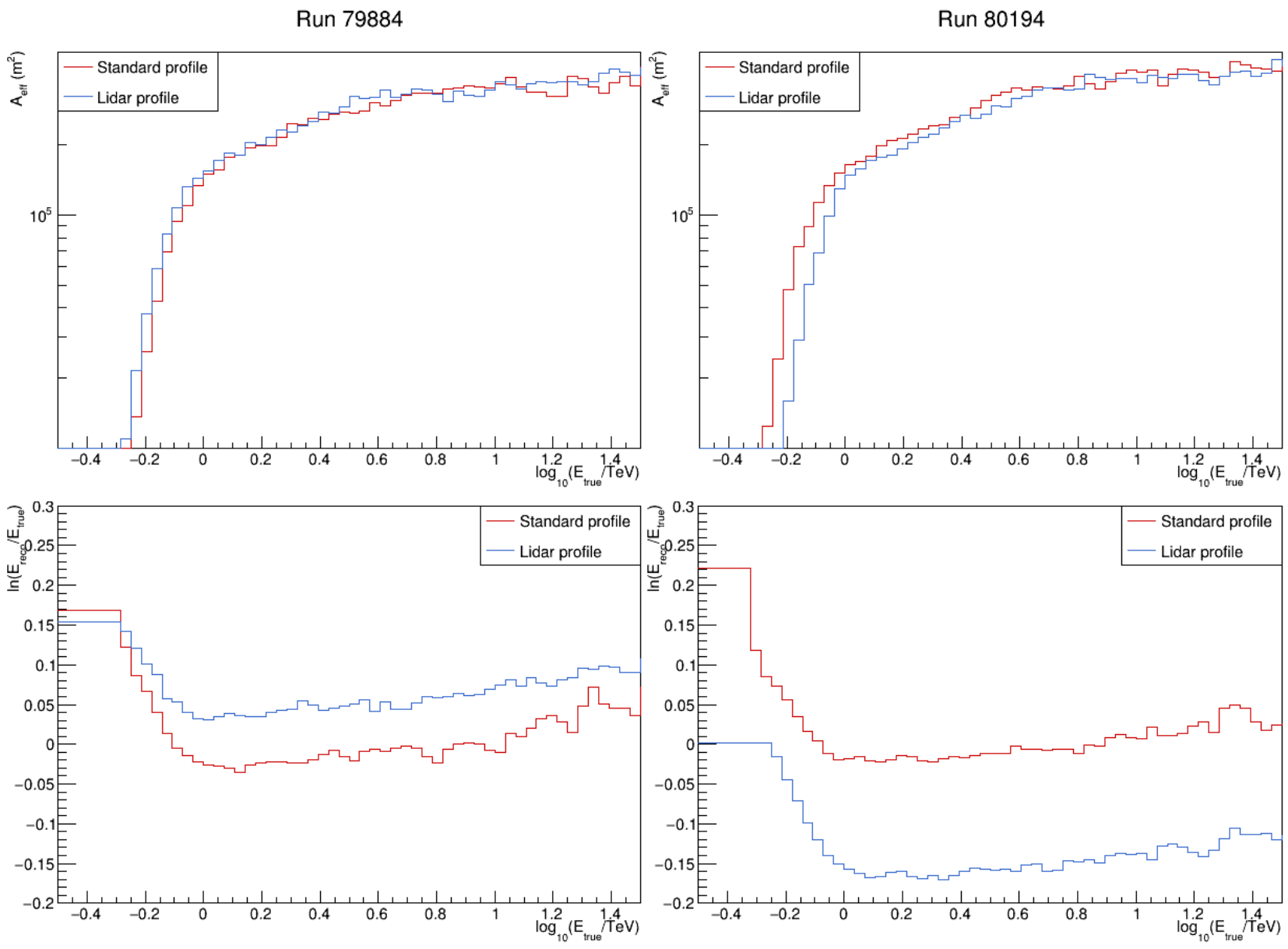

**Fig. 3.** Effective areas (top) and energy biases (bottom) obtained for the runs 79884 and 80194, simulated with the standard profile (red) and the corresponding lidar profile (blue).

For each run, we simulated $10^7$ events with a power-law spectrum of $E^{-2}$ from 0.3 TeV to 100 TeV. Each run was simulated twice: using the standard profile and the lidar profile. Figure 3 shows the IRFs obtained for the runs with the most transparent (79884) and the most opaque (80194) atmosphere in our dataset. As expected, a more opaque atmosphere leads to a higher energy threshold and a smaller reconstructed energy. At 1 TeV, the effective area variations are +4% and -13% with lidar data, respectively for the runs 79884 and 80184. The energy threshold can be defined where the effective area is equal to 10% of its maximum value. Thus, lidar data lead to a threshold energy shift of -5% (run 79884) and +17% (run 80194) compared to the model. The reconstructed energies (for $E_{\mathrm{true}} > 1$ TeV) are $E_{\mathrm{reco}} \sim 1.06 E_{\mathrm{true}}$ (run 79884) and $E_{\mathrm{reco}} \sim 0.91 E_{\mathrm{true}}$ (run 80194) with the lidar profiles while $E_{\mathrm{reco}} \sim E_{\mathrm{true}}$ with the standard profile (used to reconstruct the energy).

We then fit the spectrum with a power law between 300 GeV and 10 TeV, leaving the spectral index and the flux normalization free. We only use the four HESS I telescopes for the reconstruction. The run-by-run fit shows noticeable differences in the reconstructed spectra when using the lidar and standard profiles (up to 50% for the best-fit differential flux at 1 TeV and 20% for the best-fit spectral index for the run 80194). We then fit the spectra obtained with the 22 runs from 1 to 10 TeV, i.e. an energy range in which the spectrum is well described by a simple power law. Figure 4 shows the corresponding spectral energy distributions (SEDs). Both the spectral index and the differential flux at 1 TeV are similar using either the lidar profiles or the standard model (with a difference of 3% for the differential flux at 1 TeV), indicating that the model used by the H.E.S.S. collaboration seems to well represent the average atmospheric composition. The slightly higher value of the differential flux at 1 TeV, obtained with the lidar profile indicates a slightly more opaque atmosphere on average than that predicted by the standard model. In fact, a lower effective area combined with a lower reconstructed energy leads to a higher flux than the one obtained with the standard profile. Using this dataset as a whole, the impact of the lidar profiles on H.E.S.S. data is not significant.

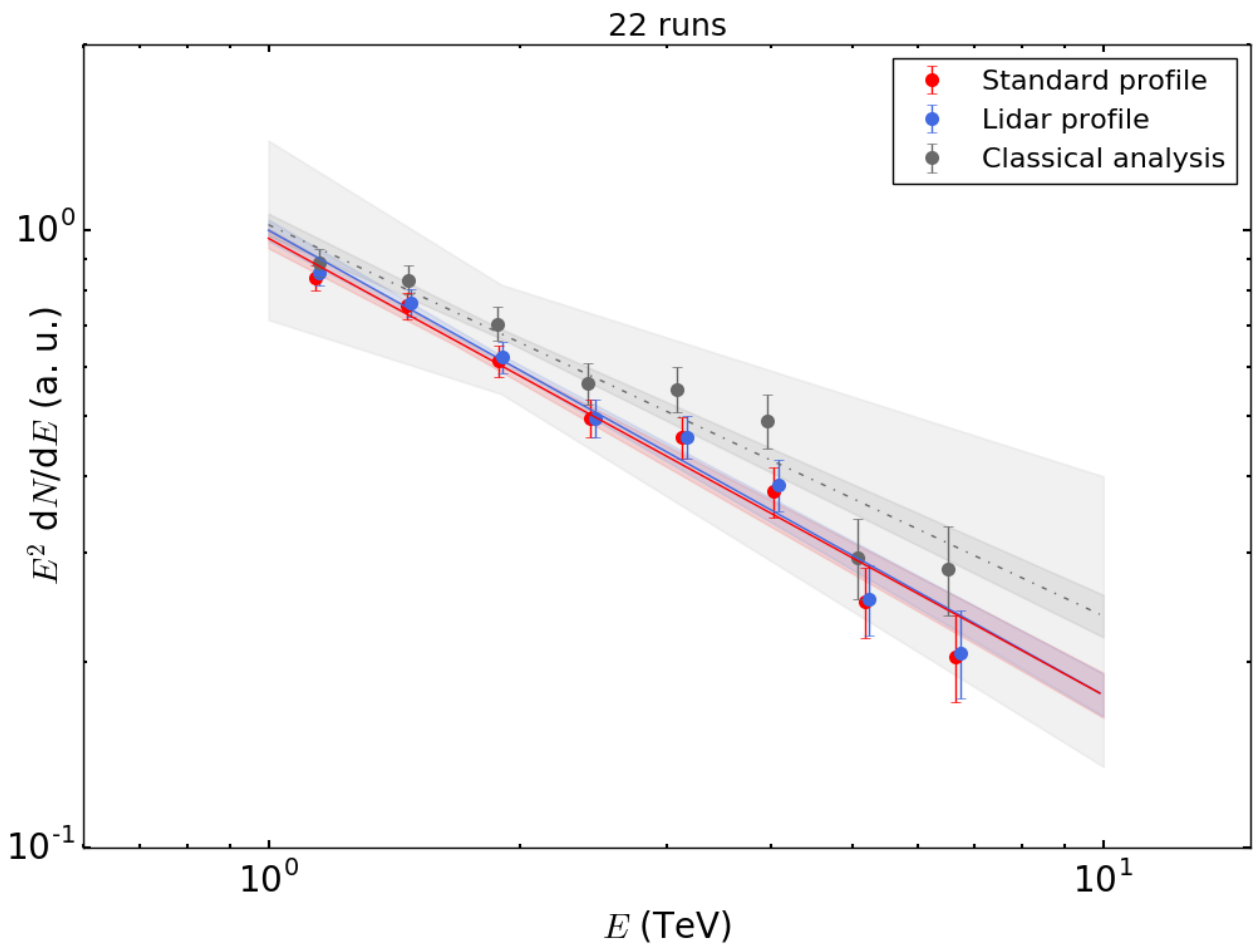

**Fig. 4.** SEDs (normalized by the SED value at 1 TeV obtained with lidar data) with 22 runs using the lidar and standard profiles. The data were fit with a power law between 1 and 10 TeV. For comparison, the SED obtained with the classical analysis is also represented, taking into account the systematic errors (± 20% on the flux normalization and ± 0.2 on the spectral index).

This result was not unexpected and is also reassuring since about half of the runs have a higher OD value than the model; thus the effect is expected to be balanced with the other half. We also reconstructed the spectrum using the classical analysis (with interpolated IRFs) and taking into account systematic uncertainties. As shown in Figure 4, the two RWS analyses (with the standard and lidar profiles) are contained in the systematic errors.

We then studied the normalization dispersion when using the lidar and standard profiles. To do so, we fit the data from each run with a power-law spectrum, fixing the spectral index to the best-fit value found when using the 22 runs (which is the same with the lidar and standard profiles). We performed the analysis from 1 to 10 TeV: in this conservative energy range, the IRFs are well described (far from the low-energy threshold) and the Crab Nebula spectrum is well approximated by a simple power law. Since the two datasets (from 2012 and from 2013) have different trigger configurations, we study the normalization dispersion separately. The values of the normalized differential flux at 1 TeV, obtained for each observation run, with the lidar and standard profiles, are given in Figure 5, for the two periods (2012 and 2013). To quantify the dispersion, we calculated the reduced $\chi^2$ (= $\chi^2$ / d.o.f., where d.o.f. is the number of degrees of freedom). The differential fluxes are normalized to the one minimizing the $\chi^2$ using lidar data (dotted blue curve). As shown in Figure 5, the spectra from the 11 first runs are not compatible with a constant flux from the Crab Nebula at more than 3 sigma, both using the lidar and standard profiles. The dispersion is not reduced when using the lidar profiles, but we note, however, that without the tenth run, the dispersion is slightly reduced with lidar data (25.1/9 compared to 31.4/9 with the standard model). This observation run is still under investigation, but there are also presumably other sources of systematic errors not yet understood (and apparently not related to the atmosphere). For the 11 last runs, the use of the standard profile gives an incompatibility with a constant flux at 3.3 sigma, while the use of the lidar profiles reduces the incompatiblity to 1.3 sigma. The normalization dispersion is thus notably reduced with lidar data for these 11 last runs, corresponding to the observations with the worst atmospheric conditions.

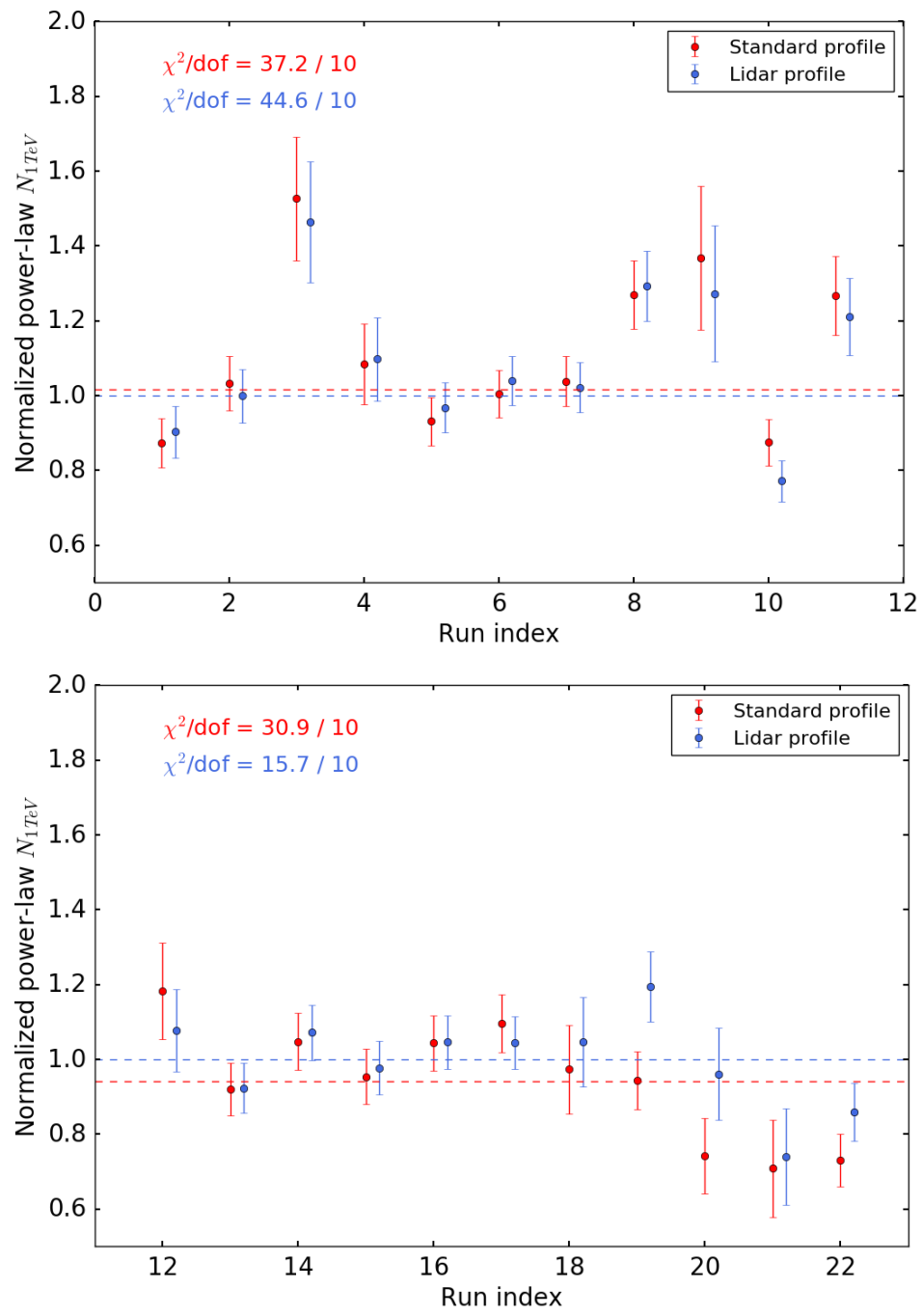

**Fig. 5.** Normalized differential flux at 1 TeV using the runs taken in 2012 (top) and in 2013 (bottom), with the standard and lidar profiles. The dotted lines represent the best-fit differential flux (minimizing the $\chi^2$).

## 4 Conclusions

We studied the impact of the lidar profiles on Crab Nebula data. Using a run-wise simulation chain, we derived the IRFs for 22 runs, with the corresponding lidar profiles and the standard profile currently used by the H.E.S.S. collaboration. We observed a notable impact on the IRFs, demonstrating the possibility to have a better knowledge of the low-energy threshold and the photon energy thanks to lidar data. Spectral differences are visible in a run-by-run analysis but are not significant when considering the 22 runs, which was not unexpected given the average OD value of these runs. We then studied the normalization dispersion dividing the dataset between runs taken in 2012 and in 2013. The normalization dispersion is notably reduced with lidar data only for the 11 last runs. This first study of the impact of the lidar profiles on H.E.S.S. data is nevertheless encouraging and more lidar data are needed to be able to better quantify how much the systematic errors can be reduced with a lidar.

*Acknowledgements*. We thank the H.E.S.S. Executive Board, for allowing us to use H.E.S.S. observation data in this technical study, and all our H.E.S.S. colleagues who facilitated the analysis of these data, particularly M. Holler and M. de Naurois.